\documentclass[aps,twocolumn,showpacs,superscriptaddress,prl]{revtex4-1}

\usepackage{graphicx}
\usepackage{dcolumn}
\usepackage{bm}
\usepackage{amssymb}

\begin{document}


\title{Dynamically generated flat-band phases in optical kagome lattices}

\author{Gia-Wei Chern}
\affiliation{Center for Nonlinear Studies and Theoretical Division, Los Alamos National Laboratory, Los Alamos, NM 87545, USA}

\author{Chih-Chun Chien}
\affiliation{Theoretical Division, Los Alamos National Laboratory, Los Alamos, NM 87545, USA}

\author{Massimiliano Di Ventra}
\affiliation{Department of Physics, University of California, San Diego, La Jolla, CA 92093, USA}


\begin{abstract}
Motivated by recent advances in the realization of complex two-dimensional optical lattices, we investigate theoretically the quantum transport of ultracold fermions in an optical kagome lattice. In particular, we focus on its extensively degenerate localized states (flat band).
By loading fermions in a partial region of the lattice and depleting the mobile atoms at the far boundary of the initially unoccupied region, we find a dynamically generated flat-band insulator, which is also a  population-inverted state. We further show that inclusion of weak repulsion leads to a dynamical stripe phase for two-component fermions in a similar setup. Finally, by preparing a topological insulating state in a partially occupied kagome lattice, we find that the topological chiral current decays but exhibits an interesting oscillating dynamics during the nonequilibrium transport.
Given the broad variety of lattice geometries supporting localized or topological states, our work suggests new possibilities to use geometrical effects and their dynamics in atomtronic devices.
\end{abstract}


\maketitle

Ultracold atoms in optical lattices have brought a great variety of quantum phenomena, including superfluid-Mott insulator transition, quantum-gas microscope, topological phases~\cite{Blochreview,GreinerNature09,Hauke12,GoldmanPNAS13,LeePRL13}, etc. The tunability of the particle density, interactions, and other properties make them a suitable test bed for studying quantum systems out of equilibrium~\cite{PolkovnikovRMP,Blochtransport,qsimulation}. Adding to the excitement is the thriving concept of atomtronics~\cite{atomtronics}, whose goal is to use ultracold atoms and optical lattices to design devices analogous to electronic ones. Recent progress in realizing different lattice geometries~\cite{Ruostekoski,becker10,struck11,panahi11,tarruell12,jo12} adds a new dimension to this rapidly growing field. In particular, the experimental demonstration of kagome optical lattices~\cite{jo12} offers a useful platform for investigating the effects of geometrical frustration on many-body phenomena.

The kagome lattice has been the archetypical setting for studying exotic magnetic phases resulting from geometrical frustration~\cite{Chalker92,Chern13,Yan11,Depenbrock12,Schnack02}. The inability of spins to simultaneously satisfy nearest-neighbor interactions leads to a highly-degenerate ground state at the classical level~\cite{Chalker92,Chern13}. The kagome lattice also features non-dispersive flat bands in the corresponding tight-binding model. In real space, the flat band is composed of a large number of degenerate localized modes, similar to the electron cyclotron states in the Laudau levels. Due to its huge degeneracy, even small interactions can potentially lead to nontrivial many-body states~\cite{Tasaki92,Wukagome,Huber10,Neupert11,Sun11,Tang11}.  The flat bands also give rise to interesting anisotropic electron transport in kagome-chain systems~\cite{Ishii06}. It is then natural to ask whether non-trivial phases can be {\it dynamically} generated in these lattices during transport of particles, and whether these phases can be observed experimentally by
taking advantage of the controllability of cold-atom systems.

Here, we make an important step towards this direction and consider nonequilibrium fermionic transport in an optical kagome lattice. In particular, we suggest a flat-band insulator by dynamically removing atoms at one edge of the lattice with a
focused laser beam, or a ``terminator beam", as that implemented in Ref.~\cite{terminator}. This insulator is a population-inversion phase that does not require any pumping to maintain it after its formation and serves as a prototypical atomtronic device. We also show that a stripe phase emerges dynamically in a similar setup consisting of two-component fermions with weak repulsive interactions.
We further study dynamics of a nontrivial topological state and its chiral current in optical kagome lattices, which could be relevant to other work on topological phases in ultra-cold atoms \cite{Hauke12,GoldmanPNAS13,LeePRL13}. Our predictions are testable and shed light on geometrical and topological effects in quantum transport.

In contrast to electronic systems with Coulomb interactions, the interactions of ultracold atoms are controllable and noninteracting fermions are readily available \cite{nonint_note1}, thus allowing us to 
isolate the lattice effects from interactions. We then first consider noninteracting fermions to clarify the intrinsic effects of the lattice geometry and include weak interactions later on. The full quantum dynamics will be monitored within
the micro-canonical formalism (MCF) \cite{Maxbook,MCFshort,TDlimit,Phase_trans} that will be modified appropriately to take into account local atomic losses as discussed below.

\begin{figure}
\includegraphics[width=1.0\columnwidth]{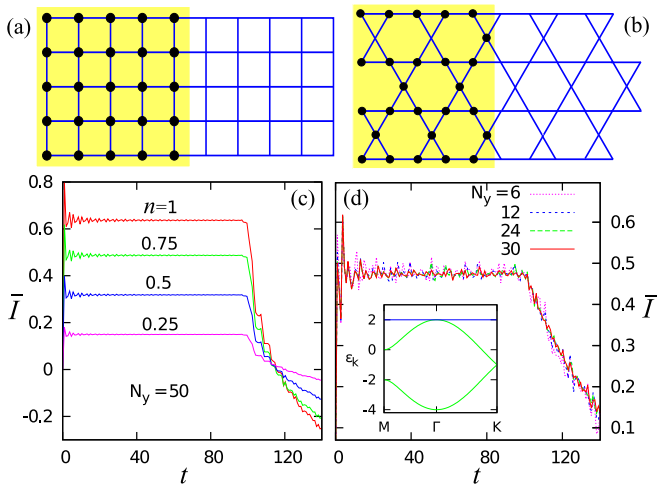}
\caption{\label{fig:lattice} (Color online) The setup (top) and the current (bottom) for (a) square and (b) kagome lattices. Initially, atoms are loaded onto the left half and flow to the right. The lattice size in the longitudinal direction is $200$ and the size in the transverse direction is $N_y$ as labeled on the plot. For the square lattice the initial fillings on the left half are $n=1,0.75,0.5,0.25$ for the curves with the highest plateau to the lowest. For the kagome lattice, the initial filling is $n=1$ and we show results with $N_y=6,12,24,30$. Inset: The band structure of the kagome lattice. The black line corresponds to the flat band.}
\end{figure}

We consider ultracold fermionic atoms loaded into a finite 2D optical lattice of square and kagome geometries as shown in Fig.~\ref{fig:lattice}(a) and (b). The square geometry serves as a yardstick for geometrical effects. We consider an isolated system described by the tight-binding Hamiltonian
$
H  = -{\bar t} \sum_{\langle ij \rangle} \bigl( c_i^\dagger c^{\;}_j + \mbox{h.c.} \bigr),
$
where ${\bar t}$ is the tunneling coefficient, $\langle ij \rangle$ denotes nearest-neighbor pairs, $c_i^\dagger$ ($c_i$)
creates (annihilates) a fermion at site $i$. The unit of time is $t_0\equiv\hbar / {\bar t}$ and is typically of a few ms \cite{MCFshort}. We first consider the case where $\bar{t}$ is uniform in the optical lattice which is possible thanks to recent progress in reducing the background lattice distortion~\cite{box_potential}. 
The system could be driven out of equilibrium by lifting an optical barrier initially blocking atoms from entering the right half~\cite{TDlimit} and we follow this protocol for 2D lattices with variable sizes in the transverse direction as shown in Fig.~\ref{fig:lattice}.
In the MCF, one computes the correlation matrix $c_{ij}(t)\equiv\langle c^{\dagger}_i(t)c_j(t)\rangle$ using $i\,dc_i/dt= [c_i,H]$. (We set $\hbar=1$.) The density $n_i(t)=\langle c^{\dagger}_{i}(t)c_{i}(t)\rangle$ and the current $I(t) = -d \langle N_L(t) \rangle / dt$,
where $N_L = \sum_{i \in L} c^{\dagger}_i c^{\;}_i$, can then be readily evaluated using $c_{ij}(t)$.  With open boundary conditions for the finite lattice, the current $I(t)$ exhibits a revival behavior when the fermions reach the far right and reflect back. Here, we focus on the physics before the revival.

{\it The role of localized modes --} Fig.~\ref{fig:lattice}(c) shows the fermionic currents in a square lattice for varying initial filling $n$. A current plateau clearly develops as time evolves, indicating the existence of a quasi-steady state current (QSSC). We have also verified that the quasi steady-state regime persists as we increase the lattice size. Our results thus provide a concrete example of QSSC in a finite 2D lattice. Similar to the QSSC found in 1D chains~\cite{MCFshort,TDlimit}, the duration of the QSSC depends linearly on the longitudinal size $N_x$ (the number of sites parallel to the current) and is independent of the initial filling. By normalizing the total current by the dimension of the transverse direction, ${\bar I} \equiv I / N_y$, we find that the magnitude of the normalized ${\bar I}$ scales linearly with the initial filling $n$, while a nonlinear dependence was observed in the 1D QSSC~\cite{MCFshort}. The difference  can be attributed to the distinct behavior of the density of states for 1D and 2D bands.

\begin{figure}
\includegraphics[width=1.0\columnwidth]{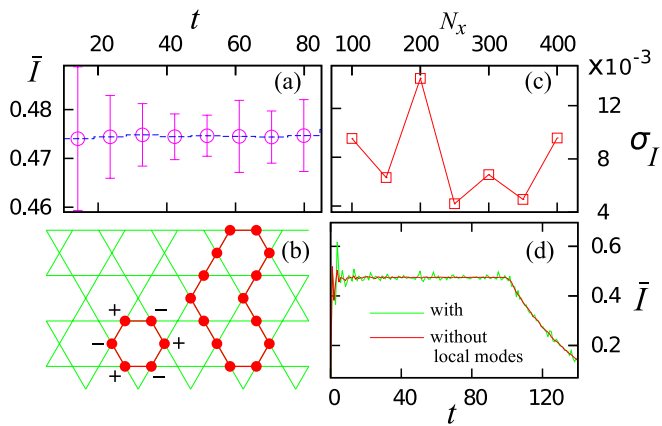}
\caption{\label{fig:fluctuations} (Color online) Fluctuations of the current in the kagome lattice. (a) The averaged current and its deviation in a window of $15t_0$ for $N_x=200$ and $N_y=30$. Inset: Two examples of localized states forming the flat band. ``$+$" and ``$-$" refer to the relative phases of the state. (b) The deviation of the current in a window of $15t_0$, taken at $t=(N_x/3)t_0$ for several values of $N_x$ with $N_y=30$ showing that the fluctuations do not decay as the system size scales. (c) The normalized current ${\bar I}$ vs. time with and without the localized modes in the MCF calculation.}
\end{figure}

We next study the current in the kagome lattice shown in Fig.~\ref{fig:lattice}(d) with an initial filling $n=1$ and varying transverse size $N_y$. In contrast to the square-lattice case, the current is now noisy and seems to fluctuate around an average value after the system is driven out of equilibrium. Fig.~\ref{fig:fluctuations}(a) shows the current averaged over 20 data points $\bar{I}(t_i)$ in a time interval $\Delta t = 15 t_0$ for a kagome lattice with $N_x = 200$ and $N_y = 30$. The error bars correspond to the standard deviations $\sigma_I = \sqrt{\langle \bar{I}^2 \rangle - \langle \bar{I} \rangle^2}$ in the same time interval. As can be seen from Fig.~\ref{fig:fluctuations}(a), the current fluctuation exhibits no discernible pattern with time. By carefully examining the dependence of the current fluctuations $\sigma_I$ on lattice dimensions, no systematic decay was observed with increasing lattice sizes; see Fig.~\ref{fig:fluctuations}(c).

These findings thus tend to support the absence of an ideal QSSC in kagome lattices, giving us a first indication that the lattice geometry significantly affects the fermionic transport via the presence or absence of localized states. In fact, our results are in line with the general theoretical study of Ref.~\cite{StefanucciPRB07} showing that a system with bound states cannot enter an ideal steady regime in the thermodynamic limit. In our case, the tight-binding Hamiltonian on the kagome lattice has two dispersive bands and a high-energy flat band at $\epsilon_{\mathbf k} = +2 {\bar t}$ as shown in the inset of Fig.~\ref{fig:lattice}(d). In real space, the basis of the flat-band modes correspond to localized states with alternating phases $\{ +1, -1, \cdots\}$ around an elementary hexagon; see Fig.~\ref{fig:fluctuations}(b). The persisting fluctuations of the current can then be attributed to the dynamic occupation of the initially empty localized states on the right half when mobile fermions of the dispersive bands flow to the right. When those localized states get partially filled, they take out some weight that was supposed to form a QSSC.
To verify this, we project out the localized modes in our calculations. By deliberately blocking all the localized states on the right from being occupied, we indeed recover a smooth QSSC; see Fig.~\ref{fig:fluctuations}(d). Other than the kagome lattice, flat bands also appear in other geometrically frustrated lattices, e.g., dice, checkerboard, and pyrochlore. Our calculation clearly demonstrates the crucial role of the flat band in destroying the QSSC.

\begin{figure}
\includegraphics[width=0.9\columnwidth]{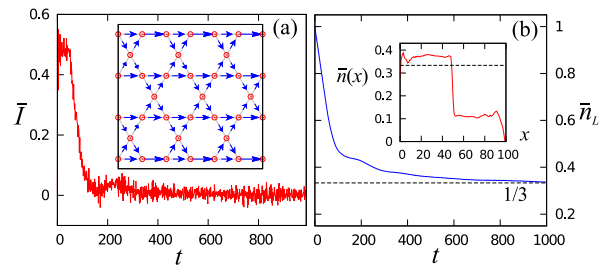}
\caption{\label{fig:device} (Color online) Dynamically generated flat-band insulator. (a) The current decays to zero. The inset shows the device, where atoms are initially loaded to the left half and a terminator beam removes particles at the far right end. (b) The density approaches the filling of the flat band, $n=1/3$. The inset shows the transversely averaged density profile at $t=250t_0$. Here $N_x=100$ and $N_y=30$.}
\end{figure}

{\it Flat-band insulator --} The initially filled localized states on the left half do not participate in the transport so their occupation number remains constant
during the dynamical process. This observation leads us to consider a way to dynamically generate a flat-band insulator. The setup is shown in the inset of Fig.~\ref{fig:device}(a). Ultracold atoms are initially loaded onto the left half with the right half of the lattice being blocked. A terminator beam similar to that used in Ref.~\cite{terminator} removes any atom that reaches the far right. In the simulations, the effect of the terminator beam is modeled by erasing all the $c_{ij}$ with $i, j$ on the boundary. Fig.~\ref{fig:device}(a) shows the time dependence of the current $I = -d\langle N_L\rangle/dt$ for such a device. The current first fluctuates around a finite value during a period that is linearly dependent on the longitudinal size of the initially filled region, reminiscent of the QSSC in the square lattice. After that it quickly decreases and remains fluctuating around zero until all dispersive atoms are depleted. The particle density $n_L(t)$ of the left half also decays to the corresponding initial filling of the flat band. In the case of one atom per site initially on the left, we have $n_L(t \to \infty) \to 1/3$, as shown in Fig.~\ref{fig:device}(b).

Importantly, the remaining atoms occupy the localized states on the left half so the system effectively becomes an insulator. The insulating nature also manifests itself in the coexistence of a vanishing current and a sharp decrease of the particle density ${\bar n}(x)$ across the middle of the system as shown in the inset of Fig.~\ref{fig:device}(b).
We emphasize that this insulating state is purely due to the initially filled flat band and no interactions between atoms are required~\cite{insulator_cmp}.
Since the flat band is at the top of the band structure for standard tunneling coefficient ${\bar t} > 0$, the above dynamical process thus also generates a state with extremely strong population inversion, a topic that has received extensive attention recently \cite{BlochInversion,MaxInversion}.
In experiments the next nearest neighbor hopping $\bar{t}_1$ may cause the flat-band insulator to decay. However, $\bar{t}_{1}/\bar{t}$ is of the order of one percent or less in typical setups \cite{Blakie04}. Thus the decay could take hundreds of $t_0$ so the flat-band insulator should be metastable and observable. An advantage of using transport to remove mobile atoms rather than depleting them using spectroscopic methods is that the terminator beam does not act on the left insulating region so there is no heating of the insulating phase.

{\it Stripe phases --} While a fully occupied flat band is nondegenerate with a uniform particle density $n = 1/3$, a partially filled flat band is a highly-degenerate many-body state. The huge degeneracy also indicates that the flat-band state is sensitive to nominally small perturbations. For example, at certain commensurate fillings of the flat band, even a small on-site repulsive interaction lifts the degeneracy and leads to a Wigner-crystalline order~\cite{Wukagome}. Here, we instead explore the possibility of lifting the flat-band degeneracy through the nonequilibrium process discussed above.

We consider fermionic atoms in two hyperfine states with weak on-site repulsion $U\sum_{i}n_{i\uparrow}n_{i\downarrow}$, where $U$ is the coupling constant. Following Ref.~\cite{TDlimit}, the equations of motion are
$
i\partial_{t} \langle c^{\dagger}_{i\sigma}c^{\,}_{j\sigma}\rangle=\bar{t}X_{\sigma}-U\langle c^{\dagger}_{i\bar{\sigma}}c^{\,}_{i\bar{\sigma}}c^{\dagger}_{i\sigma}c^{\,}_{j\sigma}\rangle+U\langle c^{\dagger}_{i\sigma}c^{\,}_{j\sigma}c^{\dagger}_{j\bar{\sigma}}c^{\,}_{j\bar{\sigma}}\rangle.
$
Here $\sigma=\uparrow,\downarrow$ represents the two components,  $X_{\sigma}\equiv \sum_{\Delta}\langle c^{\dagger}_{i+\Delta,\sigma}c^{ }_{j\sigma}\rangle-\langle c^{\dagger}_{i\sigma}c^{ }_{j+\Delta,\sigma}\rangle$, and $\Delta$ denotes the four nearest neighbors. We consider the initial condition with two atoms per site on the left half corresponding to a band insulator,  and at time $t=0$ atoms start to flow to the right half. We also impose a terminator beam at the far right end. The standard mean-field (Hartree-Fock) approximation \cite{TDlimit} of decomposing $\langle c^{\dagger}_{i\bar{\sigma}}c^{\,}_{i\bar{\sigma}}c^{\dagger}_{i\sigma}c^{\,}_{j\sigma}\rangle$ as $\langle c^{\dagger}_{i\bar{\sigma}}c^{\,}_{i\bar{\sigma}}\rangle\langle c^{\dagger}_{i\sigma}c^{\,}_{j\sigma}\rangle$ is implemented for small $U/\bar{t}$ and the two components evolve symmetrically.

\begin{figure}
\includegraphics[width=1.0\columnwidth]{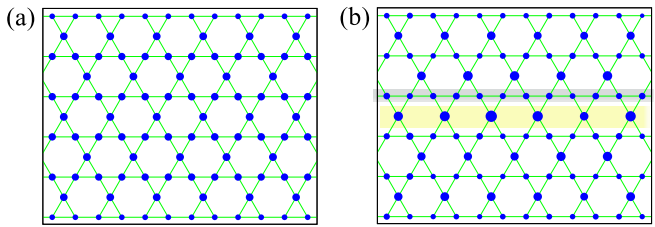}
\caption{\label{fig:density} (Color online) The density profiles of the left half for (a) noninteracting fermions in a uniform kagome lattice and (b) two-component repulsive fermions in uniform kagome lattice ($U/\bar{t}=0.5$). The radius of the circle is proportional to particle density. The shadows highlight the density modulation. These snapshots are taken for $N_x=100$ at $t = 300 t_0$.}
\end{figure}

As the system evolves, a dynamically generated stripe phase emerges on the left half lattice as shown in Fig.~\ref{fig:density}(b). A similar pattern is observed with different ratios $U/\bar{t} < 1$, where the mean-field approximation should work. With two atoms per site on the left, the initial state has a filled non-degenerate flat band and a uniform particle density. However, the degeneracy is dynamically induced due to the interaction between localized and mobile fermions during the dynamical process. This is evidenced by a slightly reduced filling fraction of the flat band during the transport process. The repulsive interaction tends to push particles toward those sites with fewer next-nearest neighbors along the current direction, giving rise to a dynamical stripe phase. We also observe a similar stripe pattern generated in noninteracting fermions by modulating the optical lattice~\cite{Schori04}.

{\it Nonequilibrium dynamics of topological currents - } The flat band in the kagome lattice touches the lower-energy dispersive band at a topologically nontrivial $\Gamma$ point ($\mathbf k = 0$)~\cite{SunPRL09}. It has been shown \cite{SunPRL09} that gapping out this band-crossing point leads to a topological insulating state with symmetry-protected gapless excitations at the boundary of the system. By preparing such a topological insulating state in the left half of the kagome lattice in our setup, one can study the nonequilibrium dynamics of the chiral currents associated with the topological phase when the atoms flow to the right.

Such a topological state can be realized with the introduction of spin-orbit coupling. It has recently been proposed that a quantum spin Hall state can be generated in the optical kagome lattice through designed laser-atom interactions to mimic the spin-orbit coupling (a gauge field)~\cite{LiuPRA10}. The effective Hamiltonian is $H_{\rm eff}=H+H_{\rm so}$, where $H$ is the tight-binding Hamiltonian described above and
$H_{\rm so}=i \lambda_{\rm so}\sum_{\langle ij\rangle} \bar{c}^{\dagger}_{i}\sigma_{z}\bar{c}_{j}+ {\rm h.c.}$ Here $\bar{c}^{\dagger}_{i}=(c^{\dagger}_{i\uparrow},c^{\dagger}_{i\downarrow})$, $\sigma_{z}$ is the diagonal Pauli matrix, and $\lambda_{\rm so}$ is the effective spin-orbit coupling constant.  The spin-orbit coupling opens a gap of order $\lambda_{\rm so}$ at the $\mathbf k = 0$ band-crossing point and endows a nonzero Chern number to the flat band (which becomes slightly dispersive).

\begin{figure}
\includegraphics[width=1.0\columnwidth]{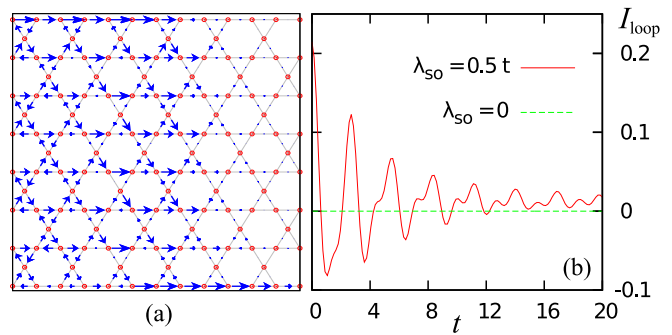}
\caption{\label{fig:c_current} (Color online) Left: A snapshot of the chiral current (the arrows) of one component  as atoms flow to the right. Right: The evolution of the total chiral current.}
\end{figure}

By initially loading two-component fermions in the left half with a filling fraction $2/3$, a quantum spin-Hall state is formed in which the two fermion species have opposite Chern numbers $C = \pm 1$. In addition, a chiral edge current is associated with each fermion component. Microscopically, the edge current results from a nonzero chiral current at each triangular loop; see Fig.~\ref{fig:c_current}(a). This chiral current may be observed as circulations in time-of-flight images~\cite{Schori04,tarruell12}. Initially, when the right-half lattice is blocked, the total chiral current on the left half has a finite value. When atoms flow to the right half, the sharp boundary separating the left and the right disappears. We followed the dynamics and found a decaying total chiral current as shown in Fig.~\ref{fig:c_current}(b). Although the total chiral current decays to zero as expected due to a terminator beam at the far right, it shows oscillations during transport. The period of the oscillation is roughly of the order of $\hbar/ {\bar t}$. This result indicates that as the system evolves, the chirality of individual fermion species reverses several times during the relaxation process.

Alternatively, a topological phase may be generated via nearest-neighbor repulsive interactions $V$ with spinless fermions~\cite{SunPRL09,LiuPRB10,WenPRB10}. By filling the particles up to the $\mathbf k = 0$ band-crossing poing, the Fermi ``surface" shrinks to a single point with a nonzero Berry connection. More importantly, this Fermi point is unstable against small perturbations~\cite{SunPRL09} and a small $V$ will gap out this Fermi point, giving rise to a quantum Hall insulator. Our setup is thus an ideal platform for investigating nonequilibrium properties of those topological phases as well.

In summary, we have shown how to utilize the localized states of a kagome lattice to design a device exhibiting a dynamically-generated flat-band insulator. Spontaneously emerging stripe phases can be observed in similar setups with or without particle interactions. Furthermore, decaying of topological states may be studied using similar protocols. The systems we suggest exploit valuable attributes of ultracold atoms and optical lattices, which may otherwise be difficult to realize in conventional solid-state systems, and broaden the possibility of utilizing geometrical or topological effects in atomtronics.

G. W. C and C. C. C. acknowledges the support of the U. S. DOE through the LANL/LDRD Program.
M. D. acknowledges support from the DOE grant DE-FG02-05ER46204.

\bibliographystyle{apsrev4-1}

\begin{thebibliography}{10}
\bibitem{nonint_note1}%
  \BibitemOpen
  \bibinfo {note} {Since $s$-wave scattering is suppressed by Pauli exclusion,
  single-component fermions are virtually noninteracting.}%
  \bibAnnoteFile{Stop}{nonint_note1}%
\bibitem{insulator_cmp}%
  \BibitemOpen
  \bibinfo {note} {The flat-band insulating state does not need any energy
  mismatch to emerge, unlike the scheme discussed in Ref.~\cite{int_ind}.}%
  \bibAnnoteFile{Stop}{insulator_cmp}%
\end{thebibliography}
%

\end{document}